\documentclass[showpacs,amsmath,amssymb]{revtex4}

\usepackage{graphicx}
\usepackage{dcolumn}% Align table columns on decimal point
\usepackage{bm}% bold math
\usepackage{amssymb}
\usepackage{amsmath}
\usepackage{latexsym}

\begin{document}
\title{Spin precession caused by spin-spin interaction between bounded electrons in quantum dots}

\author{Pei Wang\footnote{Corresponding
     author: e-mail: {\sf pwang@hbar.zju.edu.cn}}}
\author{Jin-liang Zheng}
\author{Xuean Zhao}
\address{Zhejiang Institute of Modern Physics, Zhejiang
University, Hangzhou 310027, P. R. China.}
\begin{abstract}
In this paper, we study the spin-spin interaction between two
electrons bounded in a quantum dot. The result shows that spin-spin
interaction will cause a pair of spins precessing synchronously. If
the two spins are parallel at initial time, the total spin
oscillates as a cosine function. If the two spins are antiparallel
at initial time, the total spin keeps zero. The precessing period is
proportional to the cube of quantum dot radius. For a 2D round
quantum dot with radius $R=50{\text{nm}}$, the precessing period
equals nearly $12\mu {\text{s}}$.
\end{abstract}

\maketitle

\section{Introduction}\label{1}
The spin dynamics in quantum dots attracts much attention in recent
years, because the experiments indicate long spin relaxation time
\cite{Fujisawa}\cite{Hanson1}\cite{Hanson4} in some kinds of quantum
dots. This makes the quantum dot the promising candidate for qubit,
which is the foundation of quantum computer. Many plans have been
proposed for representing qubit and building unitary gates in
quantum dots system. Some authors have researched the spin filling
in a quantum dot \cite{Hanson5} and proposed the qubit encoded by
more than one electron spin, such as encoding a qubit in singlet and
triplet states for two electrons in one dot or in two coupled dots
\cite{Hanson2}\cite{Loss}. Experiments have shown the
singlet-triplet transition can be electrically controlled
\cite{Fujisawa}\cite{Kouwenhoven}\cite{Hanson6}. Furthermore, there
are proposals for encoding a qubit in two specific spin states of
three electrons in single dot \cite{Jordan}. In order to get a good
representation of qubit and the corresponding gate sequences, it is
important to understand the interaction on spins in quantum
dots. Hyperfine interaction between the nucleus and electrons has
been considered as the most important interaction on spins in some
kinds of quantum dots \cite{Merkulov}\cite{Das1}. It has been
investigated by many authors
\cite{Erlingsson1}\cite{Erlingsson2}\cite{Khaetskii3}\cite{Kim}.

The Coulomb interaction between bounded electrons has also been
intensively studied. But according to quantum electrodynamics (QED),
Coulomb interaction is only the non-relativistic approximation of
electron-electron interaction. Considering the relative nature of
electron movement, spin-spin interaction, which can be regarded as
the interaction between two magnetons, should be included. There is
great difference between Coulomb interaction and spin-spin
interaction, because the latter can exert forces on spins.
Considering spin-spin interaction, only the total angular momentum
conserves. Spin-spin interaction has been researched in condensed
matter physics after the QED theory was discovered. Overhauser
\cite{Overhauser} in his early articles has discussed the spin
relaxation caused by spin-spin interaction between Bloch electrons.
He found that the effect was much smaller than that caused by
spin-orbit interaction and could be ignored. But in a quantum dot,
the spin-orbit coupling is efficiently suppressed, while spin-spin
interaction is enhanced because electrons are much closer to each
other. On the other hand, the experiment about spin filling shows
that there exist unpaired electrons in a multi-electron quantum dot.
Then there exists the spin-spin scattering process caused by
spin-spin interaction.

In quantum dots, spin-spin interaction does not cause spin
relaxation or dephasing. But it has non-trivial influence to the
evolution of spins in quantum dots.
Therefore it is obliged to study the effects of spin-spin interaction.

The paper is organized as follows: in Sec. II we give the
Hamiltonian of spin-spin interaction from QED theory. Then we
discuss the selection rule of this interaction in Sec. III. In Sec.
IV and Sec. V, we study the evolution process of two parallel and
antiparallel spins respectively. Sec. VI is a brief conclusion.

\section{spin-spin interaction}\label{2}

The Hamiltonian of QED theory can be expressed as
\begin{eqnarray}
H= \sum_s\int d^3\mathbf{p} \omega_\mathbf{p} c^\dag_{\mathbf{p},s}
c_{\mathbf{p},s} + \sum_\lambda \int d^3\mathbf{k}
(-g_{\lambda \lambda} |\mathbf{k}|)
a^\dag_{\mathbf{k},\lambda} a_{\mathbf{k},\lambda} -e \int d^3
\mathbf{r}\bar{\psi} \gamma^\mu \psi A_\mu.
\end{eqnarray}
Here we have omitted the terms creating or annihilating positrons.
The Fourier decompositions of the field operators are
\begin{eqnarray}\nonumber
\psi(\mathbf{r})= && \sum_s \int \frac{d ^3 \mathbf{p}}{(2 \pi) ^{3/2}}
\sqrt{\frac{m}{\omega_\mathbf{p}}} c_{\mathbf{p},s} u(\mathbf{p},s)
e^{i \mathbf{p} \mathbf{r}} ,\\
A^\mu(\mathbf{r}) = && \int \frac{d^3 \mathbf{k}}{ \sqrt{2 |\mathbf{k}|
(2\pi)^3}} \sum_\lambda (a_{\mathbf{k},\lambda}
\epsilon^\mu (\mathbf{k},\lambda) e^{i \mathbf{k} \mathbf{r}}
+a ^\dag_{\mathbf{k},\lambda}
\epsilon^\mu (\mathbf{k},\lambda) e^{- i \mathbf{k} \mathbf{r}}) .
\end{eqnarray}
By using the canonical transformation and noticing the relation
$\sum_\lambda g_{\lambda \lambda} \epsilon _\mu (\mathbf{k},\lambda)
\epsilon_\nu (\mathbf{k},\lambda) = g_{\mu \nu}$, one can get an
effective electron-electron interaction
\begin{eqnarray}\label{eff}
V_{eff}=\sum_{s, s',s'', s'''}\int d^3 \mathbf{p} d^3 \mathbf{p'} d^3 \mathbf{k}
\frac{e^2}{2 (2\pi)^3 |\mathbf{k}|^2} \bar {u}(\mathbf{p}+ \mathbf{k},s)
\gamma^\mu u(\mathbf{p}, s') \bar {u}(\mathbf{p'}- \mathbf{k},s''')
\gamma_\mu u(\mathbf{p'}, s'') c^\dag_{\mathbf{p}+ \mathbf{k},s}
c_{\mathbf{p}, s'} c^\dag_{\mathbf{p'}- \mathbf{k},s'''} c_{\mathbf{p'}, s''}.
\end{eqnarray}
Here we have used the low energy approximation $\omega_\mathbf{p}
\approx m$. There are two terms in this effective Hamiltonian which
exert forces on the spins. The first is the spin-spin interaction
which is the coupling between one electron spin and the magnetic
field caused by another electron spin. And the second is the
spin-current interaction which is the coupling between the magnetic
moment of one electron and the magnetic field produced by the
translational motion of another. The spin-current interaction
between bounded electrons can be neglected. The spin-spin
interaction can be divided into two terms
\begin{equation}
V_{s-s}  = V_D + V_{ex}.
\end{equation}
The first term $V_D$ represents flip of a pair of
parallel spins, and can be called parallel interaction. The
second term $V_{ex}$ represents flip of a pair of
antiparallel spins, and can be called antiparallel interaction.
According to Eq. (\ref{eff}), the parallel interaction can be expressed as
\begin{equation}\label{intD}
V_D  = \int {d^3 \mathbf{k} d^3 \mathbf{p} d^3 \mathbf{p}'
\frac{{e^2 \hbar ^2 \left( {k_x +
ik_y } \right)^2 }}{{8m^2 c^2 \epsilon _0(2\pi )^3 |\mathbf{k}|^2 }}c_{\mathbf{p} +
\mathbf{k}, \downarrow }^\dag  c_{\mathbf{p}' - \mathbf{k}, \downarrow }^\dag
c_{\mathbf{p}', \uparrow } c_{\mathbf{p}, \uparrow } }
 +h.c.,
\end{equation}
where $\uparrow$ and $\downarrow$ denote the spin-up and spin-down
states in $z$ direction respectively. And $c_{\mathbf{p},\uparrow}$ denotes the
annihilation operator of spin-up electrons. The antiparallel interaction
can be expressed as
\begin{equation}
V_{ex}  = \int {d^3 \mathbf{k} d^3 \mathbf{p} d^3 \mathbf{p}'
\frac{{ - e^2 \hbar ^2 \left( {|\mathbf{k}|^2  + k_z^2 }
\right)}}{{4m^2 c^2 \epsilon_0 (2\pi )^3 |\mathbf{k}|^2
}}c_{\mathbf{p} + \mathbf{k}, \uparrow }^\dag c_{\mathbf{p}' -
\mathbf{k}, \downarrow }^\dag  c_{\mathbf{p}', \uparrow }
c_{\mathbf{p}, \downarrow } }.
\end{equation}
To study the interaction between bounded electrons,  the Hamiltonian
should be expressed in real space. Eq. (\ref{intD})
can be transformed into
\begin{equation}\label{intD2}
V_D  = \int {d^3 \mathbf{k} d^3\mathbf{r}_1 d^3\mathbf{r}_2
\frac{{e^2 \hbar ^2 \left( {k_x  + ik_y } \right)^2 e^{i
\mathbf{k}\left( {\mathbf{r}_1 - \mathbf{r}_2 } \right)} }}{{8m^2
c^2 \epsilon _0 (2\pi )^3 |\mathbf{k}|^2 }}\psi _ \downarrow ^\dag
\left( {\mathbf{r}_1 } \right)\psi _\downarrow ^\dag \left(
{\mathbf{r}_2 } \right)\psi _ \uparrow  \left( {\mathbf{r}_2 }
\right)\psi _ \uparrow \left( {\mathbf{r}_1 }\right)} + h.c..
\end{equation}
The $\mathbf{k}$-integral is difficult to calculate for arbitrary
$\mathbf{r}_1$ and $\mathbf{r}_2$ in 3D space. In this work, we
study spin-spin interaction in a quasi-2D quantum dot sited in
$x$-$y$ plane. Then $\mathbf{r}_1-\mathbf{r}_2$ sites in $x$-$y$
plane at $z=0$, and $\mathbf{k}$ can be integrated out from Eq.
(\ref{intD2})
\begin{equation}
\int {d^3 \mathbf{k} \frac{{\left( {k_x  + ik_y } \right)^2 e^{i
\mathbf{k}\cdot (\mathbf{r}_1-\mathbf{r}_2)} }}{{| \mathbf{k}|^2 }}
= \frac{{ - 2\pi ^2e^{2i\arg(\mathbf{r}_1-\mathbf{r}_2) } }}{{\left|
{ \mathbf{r}_1-\mathbf{r}_2} \right|^3 }}},
\end{equation}
where $\arg(\mathbf{r}_1-\mathbf{r}_2)$ is the angle between
$\mathbf{r}_1-\mathbf{r}_2$ and $x$ axis. So the parallel
interaction can be expressed as
\begin{equation}\label{double}
V_D  = \frac{{ - e^2 \hbar ^2 }}{{32\pi m^2 c^2 \epsilon _0 }}\int
{d^3 \mathbf{r}_1 d^3 \mathbf{r}_2 \frac{{e^{2i\arg \left({
\mathbf{r}_1 - \mathbf{r}_2 } \right)} }}{{\left| {\mathbf{r}_1 -
\mathbf{r}_2 } \right|^3 }}\psi _ \downarrow ^ \dag \left(
{\mathbf{r}_1 } \right)\psi _\downarrow ^\dag \left( {\mathbf{r}_2 }
\right)\psi _ \uparrow \left( {\mathbf{r}_2 } \right)\psi _ \uparrow
\left( {\mathbf{r}_1 } \right)}+ h.c..
\end{equation}
This equation is valid only when the electrons are confined in a
very thin layer. Accordingly, antiparallel interaction can be
expressed as
\begin{eqnarray}
V_{ex} &=&-{\frac{{e^2\hbar ^2}}{{2m^2c^2\epsilon _0}}}\int
{d^3\mathbf{r}\psi _{\uparrow }^\dag (\mathbf{r}) \psi _{\downarrow
}^\dag ( \mathbf{r} ) \psi
_{\uparrow } ( \mathbf{r} ) \psi _{\downarrow } ( \mathbf{r} ) } \nonumber\\
&&-\frac{{e^2\hbar ^2}}{{16m^2c^2\epsilon _0\pi }}\int {d ^3 \mathbf{r}_1 d^3\mathbf{r}_2 \frac 1{{%
\left| {\mathbf{r}_1-\mathbf{r}_2}\right| ^3}}\psi _{\uparrow
}^\dag\left( \mathbf{r}_1\right) \psi _{\downarrow }^\dag\left(
{\mathbf{r}_2}\right) \psi _{\uparrow }\left( {\mathbf{r}_2}\right)
\psi _{\downarrow }\left( \mathbf{r}_1\right) }. \label{exchange}
\end{eqnarray}
The antiparallel interaction conserves the total spin. From this
point of view, it is similar to the exchange interaction. But there
is difference between them. The exchange interaction need the
overlap of wave functions, while antiparallel interaction need not.

\section{spin-spin scattering of bounded electrons in quantum dots}\label{3}

We consider a quasi-2D round quantum dot with radius $R$. The
Hamiltonian is
\begin{equation}\label{ha}
H=H_0+V_{s-s},
\end{equation}
where $H_0=\sum_{\lambda,\sigma} \varepsilon_{\lambda}
c_{\lambda,\sigma}^\dag c_{\lambda,\sigma}$ is the bounded energy
and $V_{s-s}$ the spin-spin interaction. Because the number of
electrons in quantum dots is definite, the Coulomb charging energy
is a constant and is omitted. We assume the confining potential is
zero inside the quantum dot and infinite outside it. It is well
known that the bounded wave function can be expressed as Bessel
functions in polar coordinates
\begin{equation}\label{wave}
\psi_{m,n} ( \rho ,\theta ) = \frac{1}{R \sqrt{\pi
}J_{|m|+1}(Z_{m,n})}J_{m} \left( {\frac{{Z_{m,n} \rho}}{R}}
\right)e^{ im \theta },
\end{equation}
where $m=0,\pm1,\cdots$ is the angular momentum in $z$ direction and
$Z_{m,n}$ the $n$th zero of Bessel function $J_{m}$. The bounded
energy is
\begin{equation}
\varepsilon_{m,n} = \frac{{\hbar ^2 }}{{2m^* R^2 }}Z_{m,n}^2.
\end{equation}
The spin-spin interaction matrix is calculated
\begin{eqnarray}
&&\langle {m_3n_3} \downarrow {,m_4n_4\downarrow
}|V_D|{m_1n_1\uparrow ,m_2n_2\uparrow }\rangle
=-\frac{{1}}{{4c^2}}\int {d^3 \mathbf{r}_1 d ^3 \mathbf{r}_2
\frac{{e^{2i\arg
\left( {\mathbf{r} _1-\mathbf{r} _2}\right) }}}{{\left| {\mathbf{r} _1-\mathbf{r}_2}\right| ^3}}\times } \nonumber\\
&&\times {[\psi _4^{*}\left( {\mathbf{r}_2}\right) \psi _3^{*}\left(
{\mathbf{r}_1}\right) \psi _1^{}\left( {\mathbf{r}_1}\right) \psi
_2^{}\left( {\mathbf{r}_2}\right) -\psi
_4^{*}\left( {\mathbf{r}_1}\right) \psi _3^{*}\left( {\mathbf{r}_2}\right) \psi _1^{}\left( {%
\mathbf{r}_1}\right) \psi _2^{}\left( {\mathbf{r}_2}\right) ]}
\label{double2},
\end{eqnarray}
where $\psi_i$ denotes the wave function $\psi_{m_i,n_i}$ in Eq.
(\ref{wave}). We work in the atomic units, where
$\hbar=e=m=4\pi\epsilon_0=1$. By substituting Eq. (\ref{wave}) into
Eq. (\ref{double2}), we obtain
\begin{equation}\label{coupling0}
\langle {m_3 n_3 \downarrow ,m_4 n_4 \downarrow }|V_D | {m_1
n_1\uparrow ,m_2 n_2 \uparrow }\rangle= \frac{ - \gamma_D}{4\pi ^2
c^2 R^3 },
\end{equation}
where $c=1.370 \times 10^2$ is the light velocity and $\gamma_D$ a
coefficient which depends only upon the quantum numbers $m$ and $n$
of initial and final states. $\gamma_D$ can be expressed as
\begin{eqnarray}
\gamma _D &=&\frac
1{J_{|m_1|+1}(Z_1)J_{|m_2|+1}(Z_2)J_{|m_3|+1}(Z_3)J_{|m_4|+1}(Z_4)}\int
_0^1 d\rho _1d\rho _2\int_0^{2\pi } d\theta _1d\theta _2 \nonumber\\
&&\frac{\rho _1\rho _2e^{2i\varphi }J_{m_1}(Z_1\rho
_1)J_{m_2}(Z_2\rho _2)e^{i(m_1\theta _1+m_2\theta _2)}}{(\rho
_1^2+\rho _2^2-2\rho _1\rho _2\cos (\theta _1-\theta
_2))^{3/2}}[J_{m_3}(Z_3\rho _1)J_{m_4}(Z_4\rho
_2)e^{-i(m_3\theta _1+m_4\theta _2)} \nonumber\\
&&-J_{m_3}(Z_3\rho _2)J_{m_4}(Z_4\rho _1)e^{-i(m_3\theta
_2+m_4\theta _1)}], \label{coefficient1}
\end{eqnarray}
where $\varphi=\arg(\rho_1 e^{i \theta _1}-\rho_2 e^{i \theta _2})$
and $Z_i=Z(m_i,n_i)$. Above integral is calculated numerically by
Monte Carlo method. The $\gamma_D$ is nonzero only when
$m_1,m_2,m_3$ and $m_4$ satisfy the relation
\begin{equation}\label{selection}
m_3  + m_4  - m_1  - m_2 = 2.
\end{equation}
This can be explained by the conservation of total angular momentum.
The spin-spin interaction satisfies the commutation relation
$[V_D,L_z+S_z]=0$. And the bounded states are the eigenstates of
$L_z$ and $S_z$. So the matrix element is not zero only when Eq.
(\ref{selection}) is satisfied. We also find that $\gamma_D$
satisfying the selection rule has the magnitude of $10^2\sim10^3$.
For example, $\gamma_D=3.823\times10^3$ when the initial and final
states are $|01\uparrow,-11\uparrow\rangle$ and
$|01\downarrow,11\downarrow\rangle$ respectively.

The second term of antiparallel interaction in Eq. (\ref{exchange})
denotes the interaction between two electrons at different position,
which is proportional to $|r_1 - r_2|^{-3}$ and decreases sharply
with the distance of two electrons increasing. The size of a quantum
dot is usually much larger than the length unit of Bohr radius
$0.529\AA$. So the second term is usually much smaller than the
first term and can be neglected. The antiparallel interaction in
atomic units is then
\begin{equation}\label{exchange2}
V_{ex}  = \int {d^3\mathbf{r} \frac{{ - 2\pi}}{{ c^2  }}\psi _
\uparrow ^ \dag \left( \mathbf{r} \right)\psi _ \downarrow ^ \dag
\left( \mathbf{r} \right)\psi _ \uparrow  \left( \mathbf{r}
\right)\psi _ \downarrow \left( \mathbf{r} \right)} .
\end{equation}
The antiparallel interaction matrix is calculated
\begin{equation}\label{coupling2}
\langle {m_3 n_3 \uparrow ,m_4 n_4 \downarrow }|V_{ex} | {m_1
n_1\uparrow ,m_2 n_2 \downarrow }\rangle =
 \int {d ^3\mathbf{r}\frac{{ - 2\pi }}{{c^2 }}\psi _3^* \left( \mathbf{r} \right)\psi
_4^* \left( \mathbf{r} \right)\psi _1 \left( \mathbf{r} \right)\psi
_2 \left( \mathbf{r} \right)} .
\end{equation}
Here we must consider the finite height of the quantum dot. The wave
function in circular cylindrical coordinates is (choosing the lowest
level in $z$ direction \cite{Erlingsson1})
\begin{equation}\label{wave2}
\psi \left( {\rho ,\theta , z} \right) = \sqrt{\frac{2}{a \pi}
}\frac{1}{R J_{|m|+1}(Z_{m,n})}J_{m} \left({\frac{{Z_{m,n} \rho
}}{R}} \right)e^{
 im \theta } \sin(\pi \frac{z}{a}),
\end{equation}
where $a$ is the height of the quantum dot. Substituting Eq.
(\ref{wave2}) into Eq. (\ref{coupling2}), we get
\begin{equation}\label{coupling3}
\langle {m_3 n_3 \uparrow ,m_4 n_4 \downarrow }|V_{ex} | {m_1
n_1\uparrow ,m_2 n_2 \downarrow }\rangle  = \frac{{ - 6\gamma_{ex}
}}{c^2 R^2 a},
\end{equation}
where $\gamma_{ex}$ is a coefficient depending only upon the quantum
numbers $m$ and $n$ of initial and final states
\begin{equation}\label{coefficient}
\gamma_{ex} = \frac{{\int_0^1 {d\rho \rho J_{m_1 } \left( {Z_1 \rho
}\right)J_{m_2 } \left( {Z_2 \rho } \right)J_{m_3 } \left( {Z_3 \rho
} \right)J_{m_4 } \left( {Z_4 \rho }\right)} }}{{J_{\left| {m_1 }
\right| + 1} \left( {Z_1 } \right)J_{\left| {m_2 } \right| + 1}
\left( {Z_2 }\right)J_{\left| {m_3 } \right| + 1} \left( {Z_3 }
\right)J_{\left| {m_4 } \right| + 1} \left( {Z_4 } \right)}}.
\end{equation}
Due to the conservation of angular momentum, the $\gamma_{ex}$ is
not zero only when next selection rule is satisfied
\begin{equation}\label{selection2}
m_1  + m_2  - m_3  - m_4 = 0.
\end{equation}
The $\gamma_{ex}$ satisfying the selection rule has the magnitude of
$10^0$. For example, $\gamma_{ex}=0.718$ when the initial and final
states are $|0 1 \uparrow, -1 1 \downarrow\rangle$ and $|-1 1
\uparrow, 0 1 \downarrow\rangle$ respectively.

\section{The evolution of two parallel spins}\label{4}

We propose a quantum dot with two parallel spins along $z$
direction. In experiments, this can be realized by successively
injecting two spins into a quantum dot with even number of electrons
\cite{Fujisawa}\cite{Hanson2}. Electrons in two highest levels are
unpaired, while the other levels are occupied by paired spins. So
one can assume there are only two electrons, occupying two nearby
levels. The spins will flip together due to parallel interaction,
when the final and initial states are degenerate and the transition
satisfies the selection rule of Eq. (\ref{selection}). If the
initial and final states are non-degenerate, the energy difference
is usually much larger than the spin-spin interaction energy. So the
non-degenerate transition rate is very small and can be ignored. We
assume the spins are up at initial time. Arbitrary initial state
$|m_1 n_1 \uparrow,m_2 n_2\uparrow\rangle$ has at most three
degenerate states $|-m_1 n_1 \downarrow,m_2 n_2\downarrow\rangle$,
$|m_1 n_1 \downarrow,-m_2 n_2\downarrow\rangle$ and $|-m_1 n_1
\downarrow,-m_2 n_2\downarrow\rangle$. They do not always satisfy
the selection rule of Eq. (\ref{selection}). Only when $|m_1|$ and
$|m_2|$ are consecutive numbers or one of them is equal to  $-1$,
there exists one final state satisfying the selection rule. Below is
a table showing the quantum numbers $m$ and $n$ of several levels
sorted by energy. From this table, we can find the initial states
permitting spin flip.
\begin{center}
\begin{tabular}{|c|c|c|c|c|c|c|c|c|c|}\hline
{level} & {1} & {2} & {3} & {4} & {5} & {6} & {7} &{8} &{9}
\\\hline {m} & {0} & {1} & {-1} & {2} & {-2} & {0} & {3} &{-3}
&{1} \\\hline {n} & {1} & {1} & {1} & {1} & {1} & {2} & {1} &{1}
&{2} \\\hline {energy($\frac{\hbar^2}{2m^* R^2}$)} & {5.783} &
{14.68} & {14.68} & {26.37} & {26.37} & {30.47} & {40.71} &{40.71}
&{49.22}
\\\hline
\end{tabular}
\end{center}
For example, the degenerate state of the initial state $|0 1
\uparrow,-1 1\uparrow\rangle$ is $|0 1 \downarrow,1
1\downarrow\rangle$. In this situation, the initial and final states
generate the complete bases. Then the Hamiltonian can be written in
a matrix
\begin{equation}
H=H_0+V_D=
 \left[ {\begin{array}{*{20}c}
   E & V_D  \\
   V_D & E  \\
\end{array}} \right],
\end{equation}
where $E=(Z_{m1,n1}^2+Z_{m1,n1}^2)/(2m^* R^2 )$ is the total bounded
energy in atomic units. And
\begin{equation}
V_D=\langle 0 1 \downarrow,1 1\downarrow |V_D  |0 1 \uparrow,-1
1\uparrow \rangle = \frac{ - \gamma_D}{4\pi ^2 c^2 R^3 }
\end{equation}
is the parallel interaction energy. The evolution matrix evaluates
\begin{equation}
U(t)=e^{-iHt}=e^{-iEt}
 \left[ {\begin{array}{*{20}c}
   \cos V_D t & -i\sin V_D t  \\
   -i\sin V_D t & \cos V_D t  \\
\end{array}} \right].
\end{equation}
The time-dependent state can be expressed as
\begin{equation}
|\Psi(t)\rangle=U(t)|\Psi(0)\rangle =e^{-iEt}\cos V_D
t|\uparrow\uparrow\rangle-ie^{-iEt}\sin V_D
t|\downarrow\downarrow\rangle,
\end{equation}
where the quantum number $m$ and $n$ are omitted. The total spin at
$z$ direction is oscillating $S_z ( t )=\cos 2V_D t$. The spins in a
quantum dot will precess even without external magnetic field or
nuclear spins. The precession frequency $V_D /\pi$ depends on the
size of the quantum dot, being inversely proportional to cube of
radius. For example, in a quantum dot with radius $R=50\text{nm}$
the precession period is $T=\pi / V_D=12.4\mu \text{s}$. When it is
much shorter than the spin relaxation time, the precession can be
observed. In GaAs/AlGaAs 2DEG quantum dots, it is impossible to
observe the precession because the spin relaxation time is only
several nanoseconds due to hyperfine interaction \cite{Merkulov}. In
a quantum dot made of the material with zero nuclear spin (such as
Si), the spin relaxation time is much longer. Then it maybe possible
to observe this precession.

A weak vertical magnetic field depresses hyperfine interaction and
increases the spin relaxation time, but it also destroys the
precession caused by parallel interaction. Because the degenerate
transition becomes the non-degenerate transition, whose rate is much
smaller, when a magnetic field is added. But if the magnetic field
is so strong that the Zeeman splitting is comparable with the level
spacing, the non-degenerate transitions may become degenerate. And
the strong field will increase the spin relaxation time. The
electron's wave function is more compact in the presence of magnetic
field, that increases the spin-spin interaction. So the influence of
a strong magnetic field to the precession is complicated and will be
studied in future.

\section{The evolution of two antiparallel spins}\label{5}

Next we discuss the precession of two antiparallel spins in the
antiparallel interaction $V_{ex}$. We propose a quantum dot with two
antiparallel spins. At initial time, the two spins occupy two nearby
levels $|m_1 n_1 \uparrow,m_2 n_2\downarrow\rangle$. There are at
most seven degenerate final states, $|-m_1 n_1 \uparrow,m_2
n_2\downarrow\rangle$, $|m_1 n_1 \uparrow,-m_2
n_2\downarrow\rangle$, $|-m_1 n_1 \uparrow,-m_2
n_2\downarrow\rangle$, $|m_1 n_1 \downarrow,m_2 n_2\uparrow\rangle$,
$|-m_1 n_1 \downarrow,m_2 n_2\uparrow\rangle$, $|m_1 n_1
\downarrow,-m_2 n_2\uparrow\rangle$, $|-m_1 n_1 \downarrow,-m_2
n_2\uparrow\rangle$. But the only one permitted by the selection
rule of Eq. (\ref{selection2}) is $|m_1 n_1 \downarrow,m_2
n_2\uparrow\rangle$. This transition is a degenerate transition for
arbitrary initial state. The Hamiltonian of electrons is
\begin{equation}
H=\left[ {\begin{array}{*{20}c}
   E & {V_{ex} }  \\
   {V_{ex} } & E  \\
\end{array}} \right],
\end{equation}
where $V_{ex}=\displaystyle\frac{ -6 \gamma_{ex}}{c^2 R^2 a }$ is
the antiparallel energy. The diagonal term of antiparallel
interaction is omitted because it is much smaller than $E$. The
evolution matrix evaluates
\begin{equation}
U(t)=e^{-iEt}
 \left[ {\begin{array}{*{20}c}
   \cos V_{ex}t & -i\sin V_{ex}t  \\
   -i\sin V_{ex}t & \cos V_{ex}t  \\
\end{array}} \right].
\end{equation}
The state at time $t$ is expressed as
\begin{equation}
|\Psi(t)\rangle=e^{-iEt}\cos
V_{ex}t|\uparrow\downarrow\rangle-ie^{-iEt}\sin
V_{ex}t|\downarrow\uparrow\rangle.
\end{equation}
The spins precess synchronously. The precession frequency
$V_{ex}/2\pi$ depends on the size of the quantum dot, being
inversely proportional to the volume of the quantum dot. For a
quantum dot with the radius of $R=50\text{nm}$ and the height of
$a=5\text{nm}$, the precession period is $T=56\mu \text{s}$ when the
initial state is $|01 \uparrow,-11 \downarrow\rangle$. A magnetic
field along $z$ direction will not destroy the precession since the
initial and final states keep degenerate.

\section{conclusion}\label{6}
We have studied spin-spin interaction between two bounded electrons
in a quantum dot. The interaction Hamiltonian includes two terms,
the parallel interaction and antiparallel interaction, which are
shown in Eq. (\ref{double}) and Eq. (\ref{exchange}) respectively.
They cause two parallel or antiparallel spins to precess
synchronously. If the two spins are parallel, and $|m_1|$ and
$|m_2|$ are consecutive numbers or one of them is $-1$, the total
spin oscillates with a period proportional to the cube of radius of
the dot. A weak magnetic field along $z$ direction will destroy this
precession. If the two spins are antiparallel and occupy different
levels, they always precess synchronously. The period is
proportional to the volume of the dot.

\section{Acknowledgement}
This work was supported by the Natural Science Foundation
10274069, 60471052 and 16225419; the Zhejiang Provincial Natural
Foundation M603193.

\end{document}